\documentclass[floatfix,twocolumn,showpacs,preprintnumbers,amsmath,amssymb,eqsecnum,aps]{revtex4}
\usepackage{graphicx}

\begin{document}
\title{Radiation interference from sources rotating 
around Schwarzschild black holes}

\author{Raissa F. P. Mendes}
\email{rfpm@ift.unesp.br}
\affiliation{Instituto de F\'\i sica Te\'orica, Universidade Estadual Paulista,
Rua Dr. Bento Teobaldo Ferraz 271, 01140-070, S\~ao Paulo, SP, Brazil}

\author{George E. A. Matsas}
\email{matsas@ift.unesp.br}
\affiliation{Instituto de F\'\i sica Te\'orica, Universidade Estadual Paulista,
Rua Dr. Bento Teobaldo Ferraz 271, 01140-070, S\~ao Paulo, SP, Brazil}

\date{\today}
\begin{abstract}
We investigate the influence of the spacetime curvature on the 
interference of the radiation emitted by an ensemble of scalar 
sources in circular motion around a Schwarzschild black hole.
We pay particular attention to the transition from the radiating 
to the non-radiating regime as the number of sources 
increases. 
\end{abstract}

\maketitle

\section{Introduction}
\label{sec:Introduction}

Astrophysical evidences for the presence of black holes in certain 
X-ray binary systems as well as in the center of many galaxies are 
mounting~\cite{Astro}. The fact that many observational data on 
black holes come from photons emitted by surrounding matter has
motivated a more detailed investigation of radiation processes 
in these non-trivial backgrounds. In the early seventies, Misner 
and collaborators championed a research program focusing on the 
synchrotron radiation emitted by fast moving charges orbiting black 
holes~\cite{Misneretal,Breuer}. Since then, related studies have 
been pursued in various situations of interest in both classical 
and quantum frameworks (see, e.g., Refs.~\cite{P95}-\cite{CdM09} 
and references therein). In all these papers, however, attention 
was restricted to the radiation emitted by individual sources. 
As we are often interested in physical situations where a 
large number of charges are present (e.g., in accretion disks of 
compact objects), a more comprehensive analysis considering 
the combined emission of various sources would be welcome. In this 
paper, we investigate the scalar radiation emitted by an ensemble 
of sources in geodesic orbits around a Schwarzschild black hole. We focus on
the effect of the spacetime curvature on the radiation interference 
and pay particular attention to the transition from the radiating 
to the non-radiating (``magnetostatic") regime as the number of 
sources increases. 

The paper is organized as follows. In section~\ref{sec:generalframework}, 
we present the quantum field theory framework in which we work 
and establish the general formulas which will be used further 
to calculate the observables of interest. In section~\ref{sec:applications}, 
we apply the previous section results to investigate the scalar radiation 
emitted by certain source configurations and discuss the corresponding 
interference effects. Section~\ref{sec:finalremaks} is dedicated to 
our final remarks. We assume natural units $c=\hbar=G=1$ (unless stated otherwise) 
and metric signature $(+ - - -)$.

\section{General Framework}
\label{sec:generalframework}

The spherically symmetric vacuum solution of Einstein equations describing 
a black hole with mass $M$ is 
\begin{equation}
ds^2 = f(r) dt^2 - f(r)^{-1} dr^2 - r^2 (d\theta^2 + \sin^2\theta d\varphi^2),
\end{equation}
where $f(r)=1-2M/r$. Now, we introduce $N$ point scalar sources in uniform 
circular motion composing a thin disk at the equatorial plane, $\theta=\pi/2$, 
with four-velocities   
\begin{equation}
u_i^\mu(\Omega_i,R_i) = (f(R_i) - R_i^2 \Omega_i^2)^{-1/2}(1,0,0,\Omega_i),
\end{equation}
where $i = 1, \ldots, N$ labels the sources, $R_i$ denotes the radial coordinates, 
and $\Omega_i$ stands for the constant angular velocities as defined by asymptotic 
static observers. The corresponding total source density is given by
$$
j(x^\nu)=\sum_{i=1}^{N}j_i(x^\nu),
$$ 
where
\begin{equation}
	j_i(x^\nu) = {\frac{q_i}{\sqrt{-g}\, u_i^0} \delta(r-R_i) \delta(\theta - \pi/2)
	\delta(\varphi - \Omega_i t - \lambda_i)}
\label{source}
\end{equation}
is normalized so that 
$$
\int{d\sigma_i j_i(x^\nu)}=q_i
$$ 
and $d\sigma_i$ is the proper 3-volume element orthogonal to $u^\mu_i$. Here 
$g\equiv\det(g_{\mu\nu})$ and $\lambda_i = {\rm const}$ defines the initial
angular positions.

We minimally couple $j(x^\mu)$ to a real massless scalar field 
$\hat{\Phi}$ according to the interaction Lagrangian density 
\begin{equation}
\mathcal{L} = \sqrt{-g} j (x^\nu ) \hat{\Phi}.
\end{equation}
The free scalar field operator $\hat \Phi$ satisfies the massless Klein-Gordon 
equation $\Box \hat \Phi = 0$ and can be expanded in terms of creation 
$a^{ \dagger}_{\alpha \omega lm}$ and annihilation $a_{\alpha \omega lm}$ operators 
as
\begin{equation}
	\hat{\Phi} = 
	\sum_{\alpha = \rightarrow}^\leftarrow
	\sum_{l = 0}^\infty
	\sum_{m = -l}^l
	\int_0^\infty d\omega 
	[u_{\alpha \omega l m}(x^\mu) a_{\alpha \omega l m}+ {\rm H.c.}],
\label{expansion}
\end{equation}
where 
$$
[a_{\alpha \omega lm}, a^{\dagger}_{\alpha' \omega' l' m'}] 
= \delta(\omega-\omega') \delta_{l l'} \delta_{m m'} \delta_{\alpha \alpha'}.
$$
Here $\omega \geq 0 $ denotes the scalar particle frequency, 
$l \geq 0$ is the azimuthal angular momentum quantum number, 
$m \in [-l , l]$, $\alpha$ labels outgoing modes ($\alpha = \rightarrow$) 
from the past white hole horizon ${\cal H}^-$ and incoming ones 
($\alpha = \leftarrow$) from the past null infinity ${\cal J}^-$. 
The positive-frequency orthonormal solutions (with respect to the 
Klein-Gordon inner product~\cite{birrel}) satisfying 
$\Box u_{\alpha \omega l m}=0$ are denoted by
\begin{equation}
	u_{\alpha \omega l m} =
	\sqrt{\frac{\omega}{\pi}} \frac{\psi^\alpha_{ \omega l}(r)}{r} 
	Y_{l m}(\theta,\varphi) e^{-i\omega t},
\end{equation}
where $Y_{l m}(\theta, \varphi)$ are the usual spherical harmonic 
functions and $\psi^\alpha_{\omega l}$ satisfy the differential 
equation
\begin{equation}
	\left [-f(r) \frac{d}{d r}
	\left(f(r)\frac{d}{d r}\right) + V_{\rm S}(r)
	\right ] \psi^\alpha_{\omega l}(r) = \omega^2 \psi^\alpha_{ \omega l} (r)
\label{difeq}
\end{equation}
with
\begin{equation}
V_{\rm S} (r) \equiv (1-2 M/r)[2 M/r^3 + l(l + 1)/r^2].
\label{V_S}
\end{equation}
For the sake of further convenience, it is useful 
to note that the scattering potential $V_{\rm M} (r)$ 
associated with Minkowski spacetime can be obtained 
from $V_{\rm S} (r)$ by vanishing the black hole mass $M$ 
in Eq.~(\ref{V_S}): 
$V_{\rm S}(r) \stackrel{M\to 0}{\longrightarrow} 
V_{\rm M} (r) \equiv l(l + 1)/r^2$.  
In contrast to the Schwarzschild case where $\psi^\alpha_{ \omega l}$
can be expressed in terms of simple functions only  in the low- 
and high-frequency regimes, the (single) set of normalizable 
radial functions in the Minkowski case can be written in terms of 
usual spherical Bessel functions as $\psi^M_{\omega l}(r)=rj_l(\omega r)$ 
for all $\omega$.

In the low-frequency regime, the leading terms of the radial functions 
in the Schwarzschild case are (up to an arbitrary phase)~\cite{HMC1998}
\begin{equation}
	\psi^{\rightarrow}_{\omega l}(r) \approx 2 r Q_l(r/M - 1)
\label{low1}
\end{equation}
and
\begin{equation}
	\psi^{\leftarrow}_{\omega l}(r) \approx
	\frac{2^{2 l}(l!)^3(M\omega)^l r P_l(r/M - 1)}{(2 l)! (2 l + 1)!},
\label{low2}
\end{equation}
where $P_l(x)$ and $Q_l(x)$ are Legendre functions. In the high-frequency 
regime, good results can be obtained using the WKB method~\cite{CdM09}:
\begin{equation}
	\psi^{\rightarrow}_{\omega l}(x) \approx
	-i \sqrt{\frac{M}{2 \omega \kappa_{\omega l}}} 
	e^{\xi_{\omega l}(x)-\Theta_{\omega l}},
\label{wkb1}
\end{equation}
\begin{equation}
	\psi^{\leftarrow}_{\omega l}(x) \approx
	-i \sqrt{\frac{M}{2 \omega \kappa_{\omega l}}} 
	e^{-\xi_{\omega l}(x)},
\label{wkb2}
\end{equation}
where 
$x \equiv r/2M + \ln(r/2M - 1)$ 
is the dimensionless tortoise radial coordinate and 
$\kappa_{\omega l} \equiv 2 M \sqrt{V_{\rm S}(x)-\omega^2} \equiv ik_{\omega l}$.
Here, we have defined
\begin{equation}
	\xi_{\omega l}(x) \equiv \int_x^{x_+}{\kappa_{\omega l}(x')dx'}
\end{equation}
and the barrier factor
\begin{equation}
	\Theta_{\omega l} \equiv \int_{x_-}^{x_+}{\kappa_{\omega l}(x)dx},
\end{equation}
where $x_-$ and $x_+$ stand for the classical turning points satisfying
$V_{S}(x_\pm) = \omega^2$. The WKB method guaranties good results as long 
as $k_{\omega l}^{-1} d (\ln k_{\omega l})/dx \ll 1$ but it is worthwhile 
to keep in mind that full numerical calculations show that this approximation 
satisfactorily captures the qualitative behavior of the radial functions 
in the low-frequency sector besides leading to excellent results
in the high-frequency regime~\cite{CdM09}. 

Our main observable of interest will be the emitted total power 
\begin{equation}
W^{(N)}  =  
\sum_{\alpha= \rightarrow}^\leftarrow 
\sum_{l = 1}^\infty 
\sum_{m = 1}^l W^{(N)}_{\alpha l m},
\label{totalpower}
\end{equation}
where
\begin{equation}
	W^{(N)}_{\alpha l m}=\int_0^\infty d \omega \, \omega |{\cal A}_{\alpha \omega l m}|^2/T,
\label{potgeneral}
\end{equation}
$T=2\pi\delta(0)$ is the total time as measured by asymptotic observers~\cite{IZ}, 
and
\begin{eqnarray}
	{\cal A}_{\alpha \omega l m} & = & 
	\langle \alpha\, \omega \, l\, m \,|i\int{d^4 x \sqrt{-g} j(x^\mu) \hat{\Phi}(x^\mu)}|0\rangle
	\nonumber
	\\
	& = & i\int{d^4 x \sqrt{-g\, } j(x^\mu) u^{*}_{\alpha \omega l m}(x^\mu)}
\end{eqnarray}
is the emission amplitude at the tree level associated with the emission 
of a scalar particle with quantum numbers $(\alpha, \omega, l, m )$ into the 
Boulware vacuum. We note that sources with $\Omega_i ={\rm const}$ 
give rise to terms  proportional to $\delta (\omega - m \Omega_i)$
in the amplitude ${\cal A}_{\alpha \omega l m}$. Thus, $W^{(N)}$ will have no 
contributions coming from $m \leq 0$ if we assume that $\Omega_i>0$. 
This is already codified in Eq.~(\ref{totalpower}).

\section{Radiation emission and interference}
\label{sec:applications}

We now apply the formalism above to analyze the radiation 
emission from an ensemble of point sources in geodesic circular 
motion around a Schwarzschild black hole. In this case, the 
corresponding angular velocities $\Omega_i$ are related 
with the radial coordinates $R_i$ by
\begin{equation}
\Omega_i = (M/R_i^3)^{1/2}.
\label{kepler}
\end{equation}
It is particularly illuminating to investigate first the radiation 
interference restricting the number of sources to two, while allowing 
for an arbitrary angular separation between them. Afterwards, we 
consider a system of $N$ sources equally spaced around an orbit, which will 
be suited to our discussion about the transition from the radiating to 
the non-radiating regime. We shall compare the results obtained in
Schwarzschild and Minkowski spacetimes in order to extract the very 
influence of the spacetime curvature in the interference phenomenon.
\subsection{Two scalar sources}

Let us introduce a two-scalar source system by writing
$j(x^\mu) = j_1(x^\mu) + j_2(x^\mu)$, 
where $j_i$ ($i=1,2$) is given by Eq.~(\ref{source}). We define  
$q_1 \equiv q$ and $q_2 = \pm q$, where the choice between $``+"$ 
and $``-"$ will simulate ``charged" and ``neutral" configurations, 
respectively. By fixing $\lambda_1 = 0$ and $\lambda_2 \equiv \lambda$, 
we have
\begin{align}
	j(x^\mu ) 
	& = \frac{q}{\sqrt{-g\,} u_1^0} \delta (r-R_1) \delta (\theta - \pi /2)
	\delta (\varphi - \Omega_1 t)
	\nonumber
	\\
	& \pm \frac{q}{\sqrt{-g\,} u_2^0} \delta (r-R_2) \delta (\theta - \pi /2)
	\delta (\varphi - \Omega_2 t - \lambda ).	
\end{align}
The emitted power $W^{(2)}_{\alpha l m}$, as given by Eq.~(\ref{potgeneral}), 
can be cast as
\begin{equation}
	W^{(2)}_{\alpha l m} = 
	 \sum_{i=1,2} 	 {^i} W^{(1)}_{\alpha l m} 
	 + \, ^{\rm int}\!W^{(2)}_{\alpha l m},
\label{Walphalm}
\end{equation}
where
\begin{eqnarray}
	{^i} W^{(1)}_{\alpha l m}
	& = & 2 q^2 m^2 \Omega_i^2 [ f(R_i) - R_i^2 \Omega_i^2 ] 
	|\psi^\alpha_{m\Omega_i \; l}(R_i)/R_i|^2 
	\nonumber
	\\
	& \times & |Y_{l m} (\pi/2 , 0)|^2
\label{W1}
\end{eqnarray}
is the power emitted by each source $i=1,2$ separately and
\begin{eqnarray}
	^{\rm int} W^{(2)}_{\alpha l m}
	& = & 
	\pm \frac{8 \pi q^2 }{R_1R_2} \sqrt{f(R_1) - R_1^2 \Omega_1^2}
	\sqrt{f(R_2) - R_2^2 \Omega_2^2}
	\nonumber
	\\
	& \times & 
	\psi^\alpha_{m \Omega_1 \, l} (R_1) 
	\psi^\alpha_{m \Omega_2 \, l} (R_2)
	|Y_{l m}(\pi/2 , 0)|^2 
	\nonumber
	\\
	& \times & m^2 \cos (m \lambda) \Omega_1 \Omega_2 \, \delta (m \Omega_2 -m\Omega_1)/T
\label{interference}
\end{eqnarray}
is the interference term associated with the two-scalar source system.	
The $\delta (m\Omega_2 - m\Omega_1)$ term 
implies that interference effects will only appear when 
$\Omega \equiv \Omega_1 = \Omega_2 $. (This is so because in this case 
there is a well-defined  phase relationship between the emitted 
waves.) Hence, for sources in  geodesic motion belonging to a thin disk, 
the interference term~(\ref{interference}) will only contribute when 
the sources share the same orbit: $R \equiv R_1 = R_2$ 
[see Eq.~(\ref{kepler})].  We restrict our attention to this case, since the 
one where the sources do not share the same orbit does not offer a major 
challenge (once contributions coming from each source are summed up 
incoherently in the total power). 
\begin{figure}[t]
\includegraphics[width=8cm]{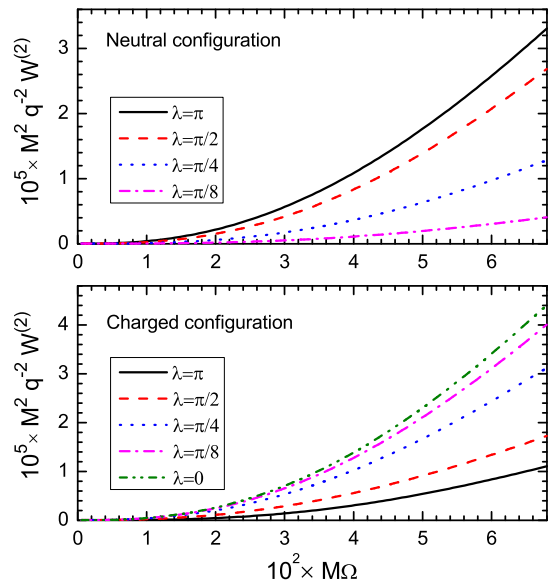}
\caption{The total emitted power $W^{(2)}$ is plotted  for stable 
circular geodesics, i.e. $M\Omega \leq 0.068$ ($R\geq 6M$), assuming 
various angular separations $\lambda$. Here, the low-frequency 
approximations~(\ref{low1})-(\ref{low2}) were used for the 
radial functions.}
\label{w2stable}
\end{figure}

\begin{figure}[t]
\includegraphics[width=7.5cm]{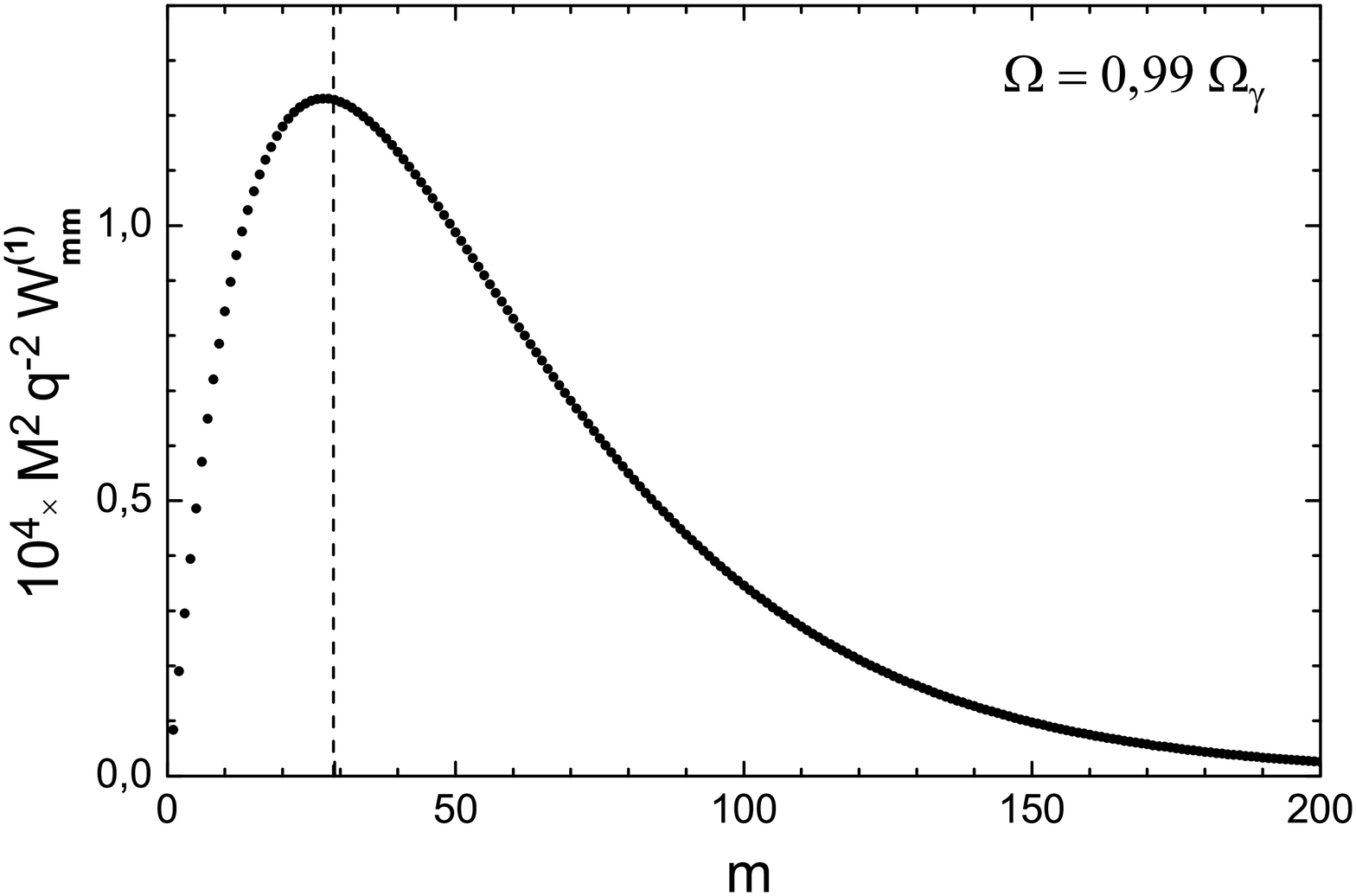}
\caption{The power $W^{(1)}_{lm}$ emitted by a single relativistic 
source with angular velocity $\Omega = 0.99 \Omega_\gamma$ is 
plotted as a function of $m$ ($l=m$). The vertical dashed line 
indicates the  $m_{\rm typ}$ value given by Eq.~(\ref{mtyp})
and illustrates the good estimation provided by this expression 
for the typical  value of the magnetic moment radiated away by sources 
at $R \gtrsim 3M$. Here, Eqs.~(\ref{wkb1})-(\ref{wkb2}) 
were used for $\psi^\alpha_{\omega l} (x)$.}
\label{w1xl}
\end{figure}

The total power~(\ref{totalpower}) associated with a two-scalar source system
where the sources at $r=R$ have angular velocities $\Omega$ and are 
separated by an angle $\lambda$  can be calculated using Eq.~(\ref{Walphalm}): 
\begin{eqnarray}
&&	W^{(2)} 	 =  
	\sum_{\alpha= \rightarrow}^\leftarrow
	\sum_{l = 1}^\infty \sum_{m = 1}^l
	4 q^2 m^2 \Omega^2 (f(R) - R^2 \Omega^2)  
	\nonumber
	\\
&& \times  
	\left| \frac{\psi^\alpha_{m\Omega \; l}(R)}{R} \right|^2
	|Y_{l m}(\pi/2 ,0)|^2 [1 \pm \cos (m\lambda )].
\label{W2tot}
\end{eqnarray}
In Fig.~\ref{w2stable}, we plot the total emitted power $W^{(2)}$ 
as a function of $\Omega$ for several
angular separations $\lambda$. For the sake of visual clarity,
we only consider sources at  stable circular geodesics, $R \geq 6M$. 
(For $3M < R < 6M$ the same monotonic increase is observed.) 
We see that for the neutral configuration, the power is the largest 
for $\lambda = \pi$ (largest dipole moment) and goes to zero as 
$\lambda$ decreases. In the  $\lambda \to 0$ limit, superposed sources, 
no radiation is emitted, as expected. On the other hand, for the 
charged configuration, the emitted power is maximum for $\lambda = 0$ 
and minimum for $\lambda = \pi$, in which case $W^{(2)}_{\alpha 11}=0$ 
and the dominant multipole contribution comes from $l=2$. 
In order to plot Fig.~\ref{w2stable}, we have used the low-frequency 
approximation given by Eqs.~(\ref{low1})-(\ref{low2}) for 
$\psi^\alpha_{\omega l}$, since particles radiated away by 
sources at $R \geq 6M$ will typically possess frequencies $\omega$ 
satisfying 
$$
\omega/M^{-1} = m \Omega_i/M^{-1} < 6.8 \times 10^{-2} \, m \ll 1.
$$
(We recall that for $R \geq 6M$, the emitted radiation is dominated 
by waves with small $m$ (see, e.g., Ref.~\cite{CHM2CQG}).) 
However, as the sources approach the innermost geodesic circular orbit, 
$R = R_{\gamma} \equiv 3M$, higher angular momentum contributions 
become increasingly important and the summations in Eq.~(\ref{totalpower}) must
be carried over larger values of $m$ and $l$. A useful relationship
estimating the typical value  of the magnetic quantum number, $m_{\rm typ}$,
carried by scalar particles radiated from relativistic sources at 
$R \gtrsim 3M$ (i.e., $\Omega \lesssim \Omega_\gamma = M^{-1}/\sqrt{27\,}$, 
where $\Omega_\gamma$ is the angular velocity of light rays at $R=R_{\gamma}$)
is provided by
\begin{equation}
	m_{\rm typ} \approx \sqrt{\frac{M}{R}\,} \frac{1-2M/R}{1-3M/R}.
\label{mtyp}
\end{equation}
In Fig.~\ref{w1xl}, we plot the power 
$$
W^{(1)}_{lm} \equiv \sum_{\alpha = \rightarrow}^\leftarrow W^{(1)}_{\alpha lm} 
$$ 
radiated by a single source 
with $M \Omega \lesssim M \Omega_\gamma \approx 0.19$, 
where [see Eq.~(\ref{W1})] 
\begin{eqnarray}
	W^{(1)}_{\alpha l m}
	& = & 2 q^2 m^2 \Omega^2 [ f(R) - R^2 \Omega^2 ] 
	|\psi^\alpha_{m\Omega \; l}(R)/R|^2 
	\nonumber
	\\
	& \times & |Y_{l m} (\pi/2 , 0)|^2
\label{W^{(1)}}
\end{eqnarray}
as a function of $m$ ($l=m$) and verify that 
Eq.~(\ref{mtyp}) is  in good agreement with the typical 
angular momentum radiated away. 
(Here, the WKB relations~(\ref{wkb1})-(\ref{wkb2}) were 
used for the radial functions $\psi^\alpha_{\omega l} (x)$.)
As $R$ approaches $3M$, $m_{\rm typ}$ grows unboundedly
and the number of relevant terms that must be considered
in the summation of Eq.~(\ref{W2tot}) increases accordingly.

\begin{figure}[t]
\includegraphics[width=8cm]{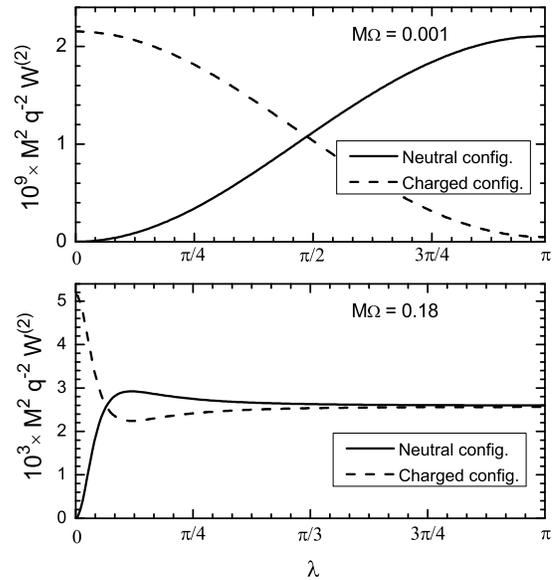}
\caption{The total power~(\ref{W2tot}) is plotted as a function 
of the angular separation $\lambda$ for neutral and charged 
configurations for small ($M \Omega = 0.001$) and large  
($M \Omega = 0.18$) angular velocities. In the small $\Omega$ 
case, the curves are roughly sinusoidal which is typical for 
dipole  emission ($l=m=1$), while in the large $\Omega$ case, 
the curves present a non-monotonic behavior because of the 
significant contribution coming from higher multipoles. Here, 
Eqs.~(\ref{low1})-(\ref{low2}) and~(\ref{wkb1})-(\ref{wkb2})
were used for $\psi^\alpha_{\omega l}$ in the top and bottom
graphs, respectively.}
\label{w2compare}
\end{figure}

In Fig.~\ref{w2compare}, we plot the total power $W^{(2)}$
as a function of $\lambda$ assuming neutral and charged 
configurations for the cases where the sources have 
small ($M \Omega \ll M \Omega_\gamma \approx 0.19$) and 
large ($M \Omega \lesssim M \Omega_\gamma \approx 0.19$) 
angular velocities.
In the small $\Omega$ case, the curves are seen to be roughly 
proportional to $1 \pm \cos \lambda$ following a dipole 
pattern~\cite{CED}. This is not so, however, in the 
large $\Omega$ case, where the various multipole contributions 
drive the curves to become nontrivial. We note, in 
particular, that interference effectively vanishes 
for large enough $\lambda$ in this case (in the sense that
the various multipole terms combine in such a way that both
sources emit as if they were independent from each other), 
driving charged and neutral configurations to behave similarly.
 
\begin{figure}[t]
\includegraphics[width=8cm]{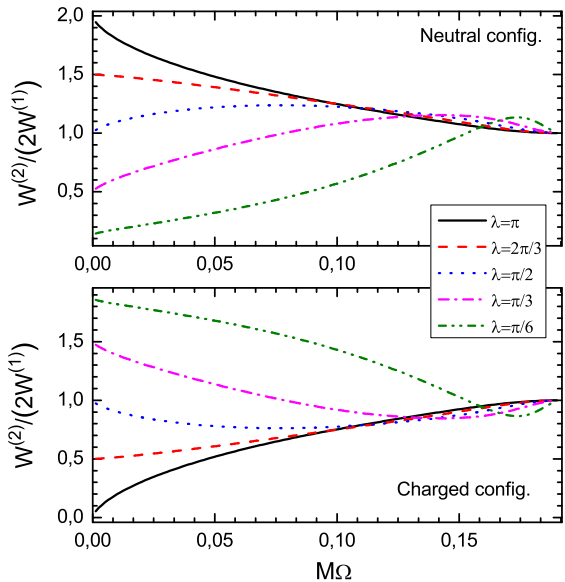}
\caption{The ratio between the total power emitted by the two-source 
system and twice the total power emitted by a single source, 
$W^{(2)}/(2 W^{(1)})$, is plotted as a function of $\Omega$ for several 
angular separations $\lambda$. As $\Omega$ approaches 
$\Omega_\gamma$ ($M\Omega_\gamma \approx 0.19$), 
this ratio goes to $1$ for every $\lambda$, implying that 
distinct sources emit irrespectively to each other for large 
enough $\Omega$. Here, Eqs.~(\ref{wkb1})-(\ref{wkb2}) were 
used for the radial functions.}
\label{w2x2w1}
\end{figure}

Interference effects are expected to be important when the typical
radiation wavelength $1/\omega_{\rm typ}$ is of order or larger 
than the distance $\lambda R $ between the sources. This leads to the 
following interference condition:
$$  
m_{\rm typ}  \lesssim [\lambda (M\Omega)^{1/3} ]^{-1},
$$ 
where we have used that $\omega_{\rm typ} = \Omega \, m_{\rm typ} $
and $R= (M/\Omega^2)^{1/3}$.
Thus, for sources with $\Omega \lesssim \Omega_\gamma$
(i.e., at $R \gtrsim 3M$), Eq.~(\ref{mtyp}) implies that we should 
not expect interference unless the sources are quite close to each other.
Fig.~\ref{w2x2w1} exhibits the  $W^{(2)}/(2W^{(1)})$ ratio as a 
function of $\Omega$ for several angular separations $\lambda$, 
where we recall that $W^{(2)}$ denotes the total power emitted by the 
two-source system as given by Eq.~(\ref{W2tot}), while $2 W^{(1)}$ 
is twice the total power emitted by a single source. (We fix $\Omega$ 
to be the same in both cases.)  Deviations from the unity indicate 
constructive ($W^{(2)}/(2W^{(1)})>1$) and destructive 
($W^{(2)}/(2 W^{(1)}) < 1$) interference. We can see from this 
graph how $W^{(2)}/(2W^{(1)})$ approaches the unity as $\Omega$ 
gets closer to $\Omega_\gamma$ and interference tends to 
effectively vanish.

\begin{figure}[t]
\includegraphics[width=8cm]{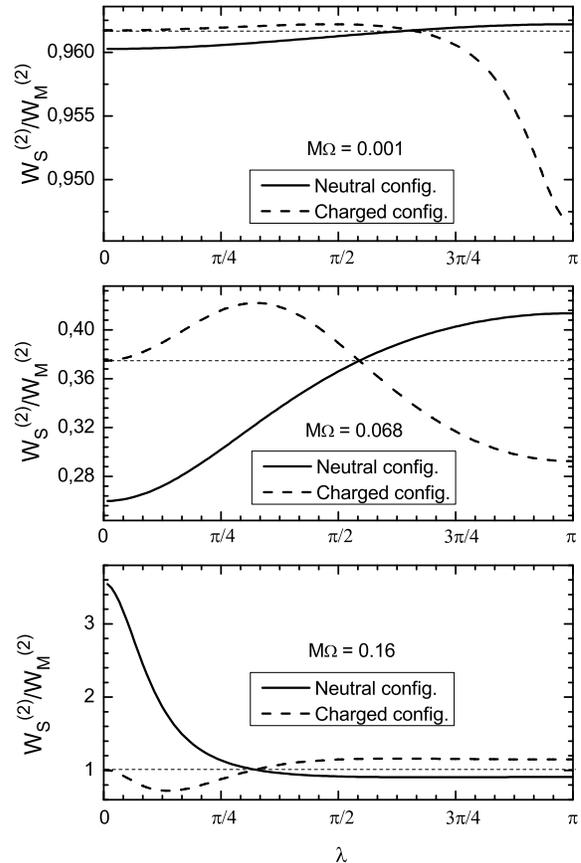}
\caption{The $W^{(2)}_{\rm S}/W^{(2)}_{\rm M}$ ratio is plotted as a 
function of the angular separation $\lambda$ for $M\Omega = 0.001, 0.68, 0.16$. 
The horizontal dashed lines correspond to $W^{(1)}_{\rm S}/W^{(1)}_{\rm M}$. 
The variation of $W^{(2)}_{\rm S}/W^{(2)}_{\rm M}$ with respect to 
$W^{(1)}_{\rm S}/W^{(1)}_{\rm M}$ indicates the influence of the 
spacetime curvature on interference and is seen to increase 
in a non-trivial way as $\Omega$ approaches $\Omega_\gamma$.}
\label{interf2}
\end{figure}

In order to investigate the role played by the spacetime curvature
on interference, we must compare $W^{(2)}/(2W^{(1)})$ as calculated
in Schwarzschild (S) and Minkowski (M) spacetimes, namely:
\begin{equation}
\frac{W_{\rm S}^{(2)}/(2W_{\rm S}^{(1)})}{W_{\rm M}^{(2)}/(2W_{\rm M}^{(1)})}
=
\frac{W_{\rm S}^{(2)}/W_{\rm M}^{(2)}}{W_{\rm S}^{(1)}/W_{\rm M}^{(1)}},
\label{curvinterf}
\end{equation}
where $W^{(1)}_{\rm M}$ is the total power emitted by a
single source as computed in Minkowski spacetime assuming 
Newtonian gravity. In this case, we relate the angular velocity 
$\Omega$ with the radial coordinate $R$ using the Kepler 
law, which formally reads as in Eq.~(\ref{kepler}) although we 
should keep in mind that in the flat spacetime case all 
quantities refer to Cartesian coordinates rather than to 
Schwarzschild ones. $W^{(2)}_{\rm M}$ is defined accordingly
for a two-source system. Rather than plotting Eq.~(\ref{curvinterf}), 
we exhibit in Fig.~\ref{interf2} $W_S^{(2)}/W_M^{(2)}$ 
and $W_S^{(1)}/W_M^{(1)}$  separately as functions of $\lambda$ for various $\Omega$ 
values assuming charged and neutral configurations. $W_S^{(1)}/W_M^{(1)}$
(dotted horizontal lines) carries information about the curvature 
influence on the radiation emitted by single sources and obviously does 
not depend on $\lambda$ , while 
$W_S^{(2)}/W_M^{(2)}$ carries information about the curvature influence
on interference as discussed in the context of Eq.~(\ref{curvinterf}).  
This is interesting to note that the term $1 \pm \cos (m\lambda)$
codifying the $\lambda$ dependence in Eq.~(\ref{W2tot}) is the same in 
Minkowski and Schwarzschild spacetimes because the angular structure of the 
corresponding metrics (which is the relevant feature here) is identical
in both cases. The nontriviality of Fig.~\ref{interf2} comes from the fact
that as one computes the {\it total} power, the sum in the quantum 
number $m$ couples the $1 \pm \cos (m\lambda)$ term to the radial functions, 
which are distinct in the curved and flat spacetimes. For low 
enough angular velocities (see $M\Omega = 0.001$), only the $l=1$ contribution 
is relevant and such coupling is ``weak". This is reflected in a small variation 
of the $W_{\rm S}^{(2)}/W_{\rm M}^{(2)}$ curve  with respect to 
$W_{\rm S}^{(1)}/W_{\rm M}^{(1)}$ (horizontal dashed line) in the whole
$\lambda$ domain (note the scale).  Now, as $\Omega$ increases 
(see $M\Omega = 0.068, 0.16$), larger values of angular momenta are excited 
and the behavior of $W_{\rm S}^{(2)}/W_{\rm M}^{(2)}$ becomes nontrivial.
Thus, despite the fact that the same $1\pm \cos (m\lambda)$ dependence 
appears in Schwarzschild and
Minkowski cases, the coupling of this term to the proper radial functions 
generates a non-trivial interplay between spacetime geometry and
interference.

\begin{figure}[b]
\includegraphics[width=8cm]{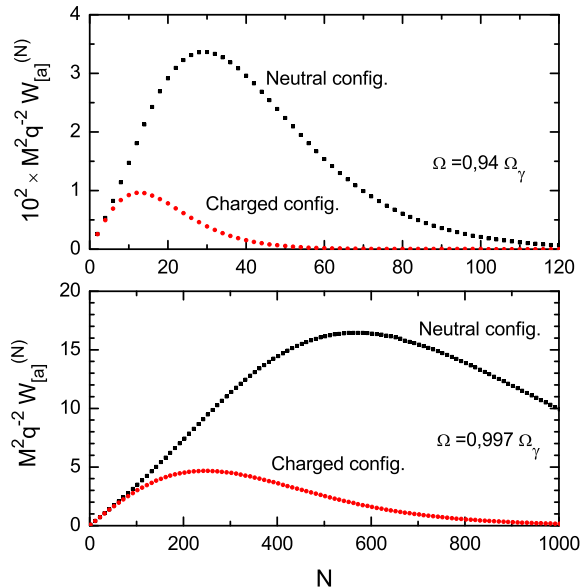}
\caption{The total power $W^{(N)}_{[a]}$ is plotted as a function of 
$N$ for two orbits in the high-frequency regime, corresponding to 
$\Omega \approx 0.94 \Omega_\gamma$ and $\Omega \approx 0.997 \Omega_\gamma$.
As $N \to \infty$, $W^{(N)}_{[a]} \to 0$ and we recover the non-radiating 
limit. Here, Eqs.~(\ref{wkb1})-(\ref{wkb2}) were used for 
$\psi^\alpha_{\omega l} (x)$.}
\label{wn}
\end{figure}

\subsection{$N$ scalar sources}
\begin{figure}[b]
\includegraphics[width=8cm]{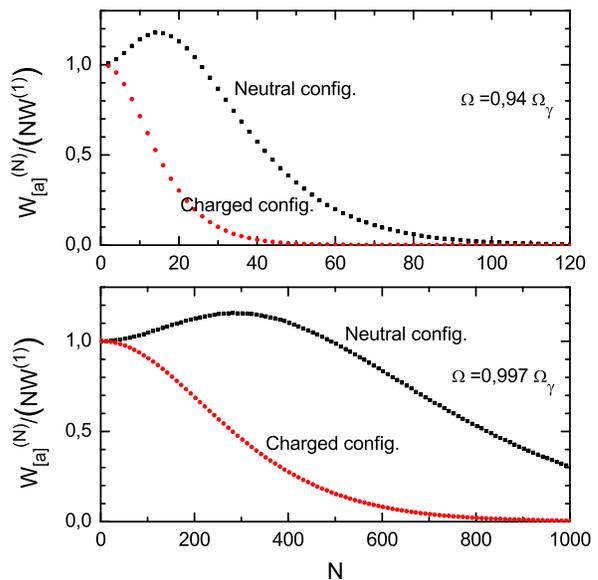}
\caption{The  $W_{[a]}^{(N)}/(N W^{(1)})$ ratio is plotted as a function 
of $N$ for $\Omega \approx 0.94 \Omega_\gamma$ and $\Omega \approx 0.997\Omega_\gamma$.
We note the existence of a region of constructive interference in the neutral
case that is not present for the charged configuration. Here, 
Eqs.~(\ref{wkb1})-(\ref{wkb2}) were used for $\psi^\alpha_{\omega l} (x)$.}
\label{wnnw1}
\end{figure}
In this section we are interested in investigating the transition 
from the radiating to the non-radiating regime as the number  of 
sources $N$ increases. They are assumed to be free and orbit the black 
hole with same angular velocity $\Omega$ ($r=R$). The $N$-source system will be 
described by
\begin{eqnarray}
	j_{[+]}(x^\mu ) & = & \sum_{n=1}^N \frac{q}{\sqrt{-g} u^0} \delta (r-R)
	\delta (\theta - \pi /2)
	\nonumber
	\\
	& \times & \delta (\varphi -\Omega t - (n-1) \lambda )
\end{eqnarray}
and
\begin{eqnarray}
	j_{[0]}(x^\mu ) & = & \sum_{n=1}^N  \frac{q (-1)^{n-1} }{\sqrt{-g} u^0} 
	\delta (r-R) \delta (\theta - \pi /2)
	\nonumber
	\\
	& \times & \delta (\varphi -\Omega t - (n-1) \lambda ).
\end{eqnarray}
for charged and neutral configurations, respectively, 
where $\lambda = 2 \pi /N$, i.e. the sources 
are equally spaced around the orbit. In the charged configuration,
all sources have the same coupling  $q$, while in the neutral one
sources with $+q$ and $-q$ appear alternately around the orbiting 
circle ($N$ is assumed even here).

The total power $W^{(N)}_{[a]}$ ($a=+,0$) emitted by these $N$-source
systems can be calculated:
\begin{equation}
	W_{[a]}^{(N)} (\Omega) = 
	\sum_{\alpha = \rightarrow}^\leftarrow 
	\sum_{l=1}^\infty \sum_{m=1}^l 
	W^{(1)}_{\alpha l m} (\Omega ) g_{[a] m}^{(N)},
\label{WN}
\end{equation}
where $W^{(1)}_{\alpha l m}$ is given in Eq.~(\ref{W^{(1)}})
and the interference factor can be cast as 
\begin{equation}
	g^{(N)}_{[+] m} = \left\{ 
	\begin{array}{lllll}
		N^2 &  {\rm when} & m = k N & {\rm for \, some}  &  k\in\mathbb{N}\\
		0 & {\rm when} & m \neq k N & {\rm for \, every} & k\in\mathbb{N}
	\end{array}\right.
\end{equation}
and
\begin{equation}
	g^{(N)}_{[0] m} = \left\{ 
	\begin{array}{lllll}
	 N^2 &  {\rm when} & m = (2k-1)N/2 & {\rm for \, some}  &  k\in\mathbb{N}\\
	 0 & {\rm when} & m \neq (2k-1)N/2 & {\rm for \, every} & k\in\mathbb{N}	
	\end{array}\right.
\end{equation}
in the charged and neutral cases, respectively.
As discussed in Fig.~\ref{w1xl}, the largest values acquired by
$W^{(1)}_{lm} \equiv \sum_\alpha W^{(1)}_{\alpha lm}$ 
happen for $l=m \approx m_{\rm typ}$ [see Eq.~(\ref{mtyp})].
The role played by $g^{(N)}_{[a] m}$ 
is twofold: on one hand, it increases the
power by a factor of $N^2$ and, on the other one, it restricts the sum in
Eq.~(\ref{WN}) to just a few significant terms: $m = N, 2N,...$ for
the charged configuration and $m = N/2, 3N/2, ...$ for the neutral one.
Thus, for low angular velocities (in which case only small multipoles 
are relevant) the emitted total power goes quickly to zero as 
the number of sources $N$ increases. Notwithstanding, if 
$\Omega \lesssim \Omega_\gamma$, higher multipoles get important 
and the power radiated by a larger number of sources can be significant. 
In Fig.~\ref{wn}, we show the total emitted power $W^{(N)}_{[a]}$ as a 
function of $N$ for two angular velocities in the high-frequency regime
assuming neutral and charged configurations. 
We note that, as the orbit approaches the innermost circular geodesic, 
the number of sources which maximize the emitted power increases as
well as the range of $N$ for which emission is significant. We also note 
that, for a given $N$, the neutral configuration radiates typically more 
than the charged one. For large enough $N$, however, interference takes 
control damping the emitted radiation.  

In Fig.~\ref{wnnw1}, we compare the total emitted power $W^{(N)}_{[a]}$ 
with $N$ times the total power emitted by a single source $W^{(1)}$ for two 
angular velocities in the high-frequency regime assuming 
neutral and charged configurations by plotting $W^{(N)}_{[a]}/(N W^{(1)})$.
For small enough $N$, we see that 
$W^{(N)}_{[a]}/(N W^{(1)}) \approx 1$, indicating that the contribution 
coming from each source is mostly independent from each other. As 
$N \to \infty$, however, interference becomes important and we recover 
the non-radiating regime. It is 
also interesting to note from Fig.~\ref{wnnw1} the existence of a region 
in which interference is constructive ($W^{(N)}/(N W^{(1)})>1$) for the 
neutral configuration in contrast to the charged case. 

The fact that interference effects should become important when the wavelengths of 
the emitted waves (of the order $1/\omega_{\rm typ}$) are larger 
than the distance $2\pi R/N$ between the sources allows us to estimate 
the number $\bar{N}$ above which interference cannot be neglected:
\begin{equation}
	\bar{N} \approx 2 \pi  \frac{M(1-2M/R)}{R(1-3M/R)}. 
\end{equation}
For the angular velocities considered in Figs.~\ref{wn} and~\ref{wnnw1},
$\bar{N} = 17$ and $350$ for $ \Omega = 0.94 \, \Omega_\gamma $ and 
$ \Omega = 0.997 \, \Omega_\gamma $, respectively.

\section{Final Remarks}
\label{sec:finalremaks}

We have considered the radiation emitted by an ensemble of scalar sources 
in circular geodesic orbits around a Schwarzschild black hole. Interference
associated with the emitted radiation was investigated and particular 
attention was given to the role played by the spacetime curvature. The 
non-radiating regime is recovered as the number  of sources $N$
increases. Finally, a simple useful result can be derived  
as follows: the power emitted by a scalar source following 
ultrarelativistic circular orbits in Schwarzschild spacetime can be 
estimated to be
\begin{equation}
	W_\delta^{(1)} \approx  10^{-4}  (q /M)^2 \delta^{-1},
\label{B}
\end{equation}
when $R \equiv (3+\delta)M$ and $0<\delta \ll 1$. As a consequence, 
we get that the total emitted power associated with $N$ sources 
will be $W^{(N)}_{[a]} \approx N W_{\delta}^{(1)}$  for $N\ll \bar{N}$ 
and $W^{(N)}_{[a]} \approx 0$ for $N\gg\bar{N}$.

The radiation component coming from the {\em collective} rotational motion 
of charges in highly ionized accretion disks orbiting black holes should 
be mostly damped by interference effects as the disk lies in 
the $R>6M$ region. However, in the short time interval which charges spend 
at $R<6M$, the role played by interference should decrease as highly 
energetic radiation is emitted, in accordance with our results. 

\acknowledgments

The authors  acknowledge full (RM) and partial (GM) financial 
support from Funda\c{c}\~ao de Amparo \`a Pesquisa do 
Estado de S\~ao Paulo (FAPESP). This work was in part inspired 
by conversations with R. Opher to whom we are grateful.

\end{document}